\documentclass[12pt]{article}
\usepackage{pic02}
\usepackage{hyperref}
\usepackage{url}
\usepackage{graphicx}

\newcommand{\bbox}[1]{\mbox{\boldmath $#1$}}
\newcommand{\case}[2]{\ensuremath{{\textstyle\frac{#1}{#2}}}}

\begin{document}

\title{\bf PROGRESS IN LATTICE QCD}
\author{
Andreas S. Kronfeld \\
{\em Theoretical Physics Department, Fermi National Accelerator Laboratory,}\\
{\em Batavia, Illinois, USA}}
\maketitle

%
% photograph of author
%  This is where we will insert a photograph. To see what it would look like,
%  uncomment the following lines.
%
%\begin{figure}[h]
%\begin{center}
%
% include photograph for proceeding version
%
%\includegraphics[height=4.5cm]{kronfeld.eps}
%
% insert a fixed vertical spacing instead for the ArXiv preprint
%
\vspace{4.5cm}
%
%\end{center}
%\end{figure}

\baselineskip=14.5pt
\begin{abstract}
After reviewing some of the mathematical foundations and numerical
difficulties facing lattice QCD, I review the status of several
calculations relevant to experimental high-energy physics.
The topics considered are moments of structure functions, which may
prove relevant to search for new phenomena at the LHC, and several
aspects of flavor physics, which are relevant to understanding $CP$
and flavor violation.
\end{abstract}
\newpage

\baselineskip=17pt

\section{Introduction}

Several areas of research in elementary particle physics (as well
as nuclear physics and astrophysics) require information about the
long-distance nature of quantum chromodynamics (QCD).
Sometimes this information can be gleaned from experiment, but often 
what one needs, in practice, are \emph{ab initio} calculations of the 
properties of hadrons.
In some cases one aims for a detailed understanding of QCD in its own
right.
In others, one simply requires a reliable calculation of hadronic
properties, so that one can study electroweak interactions or new 
phenomena at short distances.

Mathematical physicists tell us that the best way to define gauge 
theories, including QCD, is to start with a space-time lattice.
The spacing between sites is usually called $a$.
If the finite grid has $N_S^3\times N_4$ sites, then one has a finite
box size, $L=N_Sa$, and finite extent in time, $L_4=N_4a$.
Quarks are described by lattice fermion fields located at the sites, 
denoted~$\psi(x)$; 
gluons are described by lattice gauge fields located on the 
\emph{links} from $x$ to $x+a\hat{\mu}$, denoted~$U_\mu(x)$.
The key advantage to the lattice is that local gauge invariance is
simple.
The fields transform~as
\begin{eqnarray}
	\psi(x)  & \mapsto & g(x)\psi(x), \quad
		\bar{\psi}(x) \mapsto \bar{\psi}(x)g^{-1}(x), \\
	U_\mu(x) & \mapsto & g(x)U_\mu(x) g^{-1}(x+a\hat{\mu}),
\end{eqnarray}
so it is easy to devise gauge invariant actions,
\emph{i.e.}, independent of $g(x)$~\cite{Wilson:1974sk}.
If one imagines a smooth underlying gauge potential~$A_\mu(x)$ (as is 
used in continuum QCD), the relation to the lattice gauge field is
\begin{equation}
	U_\mu(x) = {\sf P}\exp\int_0^a ds\,A_\mu(x+s\hat{\mu}).
\end{equation}
Continuum QCD is defined from lattice QCD by taking $a\to0$ with $L$ 
and hadron masses fixed.
Then one takes the infinite volume limit, $L\to\infty$.
If one is interested in the chiral limit, $m_q\to0$, it 
should be taken last.
These limits are nothing radical: the lattice provides an ultraviolet
cutoff, and the finite volume an infrared cutoff.

The existence of these limits has not been proven rigorously, but, 
because of asymptotic freedom, there is not much doubt 
that this procedure works.
(If not, why does QCD work at all?)
The lattice formulation makes field theory mathematically similar to
statistical mechanics and, consequently, provides new tools.
For example, a finite lattice makes it possible to integrate the
functional integral by Monte Carlo methods.
The expectation value of observable~$\Phi$ may be written 
\begin{equation}
	\langle\Phi(\phi)\rangle = \frac{1}{Z} \int{\cal D}\phi \,
		\Phi(\phi) e^{-S(\phi)} \simeq
		\frac{1}{Z}\sum_i \Phi(\phi^{(i)}) w(\phi^{(i)})
\end{equation}
where $Z$ is chosen so that $\langle1\rangle=1$, and $\phi$ is an 
abbreviation for all fields, $\psi$, $\bar\psi$, and $U$.
The right-most expression is a numerical approximation,
with the sum running over some set of field configurations.
This numerical technique, though only one facet of lattice gauge theory,
is what most particle physicists mean by ``lattice QCD.''
So, this talk is about tools need to make numerical calculations more 
reliable, and the progress being made in calculations needed to 
interpret ``physics in collision.''

It is not so easy to descend from the mathematical high ground down
to realistic, practical numerical calculations.
Difficulties arise because QCD is a multi-scale problem.
Nature has not only the characteristic scale of QCD, 
$\Lambda_{\rm QCD}$, but also a wide range of quark masses,
leading to a hierarchy
\begin{equation}
	m_q \ll \Lambda_{\rm QCD} \ll m_Q.
\end{equation}
As a dynamical scale, rather than a parameter, there is a range 
for the QCD scale.
Some benchmarks include the scale in the running coupling 
$\Lambda_{\overline{\rm MS}}\approx250$~MeV, typical hadron masses 
like $m_\rho=770$~MeV, and the scale of chiral symmetry breaking 
$m_K^2/m_s=2500$~MeV.
The strange quark, with $m_s\approx100$~MeV, is light, and the up and 
down quarks, with $\hat{m}=\case{1}{2}(m_d+m_u)=m_s/24$ and
$m_d>m_u>0$, are especially light.
The bottom quark, with $m_b=4.25$~GeV, is heavy, and the top quark, 
with $m_t=175$~GeV, is especially heavy.
One can argue whether the charmed quark, with $2m_c=2.5$~GeV, is heavy 
or not.

Cutoffs are needed to put the problem on a computer, and they introduce
two more scales.
The idealized hierarchy is now
\begin{equation}
	L^{-1} \ll m_q \ll \Lambda_{\rm QCD} \ll m_Q \ll a^{-1}.
\end{equation}
It is impractical to expect a huge separation of scales in 
computational physics.
To explain why, some simple scaling laws are helpful.
The memory required grows like $N_S^3N_4=L^3L_4/a^4$, and these large 
exponents come because we live in $3+1$ space-time dimensions.
The CPU time needed to update gauge fields in the Monte Carlo scales 
like $a^{-(4+z)}$, where $z=1$~or~$2$, and the 4 again comes from the
dimension of space-time.
The CPU time needed to compute quark propagators scales 
like $m_q^{-p}$ where $p=1$--$3$.
The exponents $z$ and $p$ are non-zero because of properties of our 
numerical algorithms, and it seems difficult to reduce them.

The first consequence of these scaling laws is that $a^{-1}$ can 
be larger, but not much larger, than~$\Lambda_{\rm QCD}$.
Similarly, $m_q$ and $L^{-1}$ can be smaller, but not much smaller.
A~more important consequence is that improved methodology 
pays off enormously.
For example, $B$ physics with $a^{-1}\sim3\Lambda_{\rm QCD}$ instead of
$3m_b$ saves a factor $(m_b/\Lambda_{\rm QCD})^6>2^{12}$ in computing.
Such improvements are not attained by CPU power, but through new ideas.
Moreover, in computational physics the need for creativity means that 
computing facilities must be flexible, not just~big.

So, we see that finite computer resources force us to consider the 
hierarchy
\begin{equation}
	L^{-1} < m_q < \Lambda_{\rm QCD} \ll m_Q \sim a^{-1},
	\label{eq:practical}
\end{equation}
instead of the idealized one.
One should emphasize that we know how to get from the practical
hierarchy~(\ref{eq:practical}) to physical results.
The central idea is to let the computer work on dynamics at the scale
$\Lambda_{\rm QCD}$, and to use effective field theories to get the
rest. %, as sketched in Fig.~\ref{fig:eft}.
The computer runs, by necessity, with finite cutoffs and artificial
quark masses.
With effective field theories, one can strip off the artifice,
and replace it with the real world.
In doing so, one introduces theoretical uncertainties, but effective
field theories control the error analysis.

A more thorough exposition of this line of thinking can be found
in a recent review~\cite{Kronfeld:2002pi}.
Here let us emphasize the role of effective field theories by listing
some of the big ideas in lattice of the last several years:
\begin{itemize}
\item
static limit and lattice NRQCD to treat heavy quarks
\item
understanding lattice perturbation theory
(to match at short distances)
\item
non-perturbative implementation of the Symanzik effective field theory
\item
continuum HQET to control heavy-quark discretization effects
\item
novel applications of chiral perturbation theory 
\item
understanding chiral symmetry in lattice gauge theory
\end{itemize}
All but the last explicitly bring in effective field theories,
and it resonates with the usage of chiral perturbation theory to
extrapolate light quark masses.

The exception to the rule of effective field theory is something 
called the quenched approximation.
Quenched QCD is a model, so the associated uncertainty is difficult to
estimate.
For example, unquenched (usually $n_f=2$ not~3) calculations of hadron 
masses, decay constants, etc., suggest changes of 0--20\%.
Fortunately, the quenched approximation is going away.
Within a few years, I imagine that quenched calculations will no 
longer play an important role in our thinking about non-perturbative 
QCD.

The rest of this paper looks at some calculations needed to interpret 
current and future experiments.
Moments of structure functions can help obtain better parton densities 
and, hence, better predictions of cross sections at the Tevatron and 
LHC.
Lattice calculations of these moments are covered in 
Sec.~\ref{sec:moments}.
In flavor physics, the focus of many experiments, there is a pressing 
need for the hadronic matrix elements needed in leptonic and 
semi-leptonic decays, and neutral meson mixing.
There are many of these, and Sec.~\ref{sec:flavor} covers only a subset:
form factors ${\cal F}_{B\to D^*}(1)$ and $f_+(E)$ to obtain 
$|V_{cb}|$ and $|V_{ub}|$ from $B\to D^*l\nu$ and $B\to\pi l\nu$;
and matrix elements for neutral $B$, $B_s$, and $K$ oscillations,
need to constrain~$V_{td}$.
Section~\ref{sec:end} concludes with a summary and some thoughts about
the future.
	
\section{Moments of Structure Functions}
\label{sec:moments}

\begin{figure}[b!p]
	{\small (a)}\includegraphics[height=1.75in]{fcartoon.eps} \hfill
	{\small (b)}\includegraphics[height=1.75in]{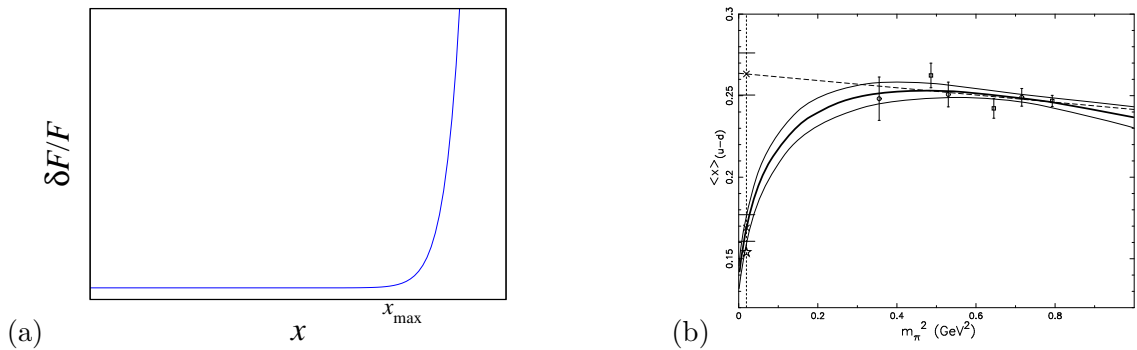}
	\caption[Deeply inelastic scattering]{\it (a) Sketch of the 
	uncertainty in measured structure functions, as a function of $x$.
	(b) Comparison of chiral extrapolations for $\langle x\rangle_{u-d}$;
	from Ref.~\cite{Dolgov:2002mn}}
	\label{fig:pdf}
\end{figure}

The rate for deeply inelastic $lp$ scattering (DIS) depends on several
structure functions, which we shall generically denote~$F(x)$.
$x$ is the momentum fraction of the struck parton, $0\le x\le1$.
In perturbative QCD, the $F$s can be related to process-independent 
parton densities, which are used to predict cross sections for 
$p\bar{p}$ and $pp$ collisions.
The DIS data peter out for $x>x_{\mathrm{max}}\sim0.7$, so, as sketched
in Fig.~\ref{fig:pdf}(a), the uncertainty explodes for the highest
values of~$x$.
High~$x$ partons are needed to produce high-mass particles.
Modern methods for parton densities directly reflect the lack of 
information~\cite{Giele:2001mr}.
They need independent (but QCD-based) knowledge of the moments to 
constrain~$F(x)$ for $x>x_{\rm max}$.

The operator product expansion relates the moments to local operators,
\begin{equation}
	\int_0^1dx\,x^{n-1}F(x,Q^2) = C_n(Q^2/\mu^2)\;
		\langle p|{\cal O}_n|p\rangle(\mu),
\end{equation}
where $C_n$ is a short-distance Wilson coefficient that is calculated 
in perturbation theory.
The matrix element on the right-hand side should be ``easy'' to
calculate: just find a lattice operator ${O_n}_{\rm lat}$ with the
same quantum numbers as ${\cal O}_n$ and calculate the matrix element
at several different lattice spacings.
We have a detailed description of lattice-spacing effects, namely,
\begin{eqnarray}
	\langle p|{O_n}_{\rm lat}|p\rangle(a) & = &
		Z^{-1}_{nm}(\mu a) \langle p|{\cal O}_m|p\rangle(\mu) +
		aK_{nj} \langle p|{\cal O}'_j|p\rangle \nonumber \\
		& + &
		aK_{\sigma F}Z^{-1}_{nm}\int d^4x\;
			\langle p|T{\cal O}_m \bar{q}\sigma F q(x)|p\rangle + O(a^2),
\end{eqnarray}
based on an effective field theory introduced by Symanzik two decades
ago~\cite{Symanzik:1979ph}.
Differences between lattice gauge theory, with $a\neq 0$, and 
continuum QCD arise at short distances comparable to~$a$.
They are lumped into short-distance coefficients $Z_{nm}$, $aK_{nj}$, 
and $aK_{\sigma F}$.
The operators on the right-hand side are, in the Symanzik effective 
field theory, defined with a continuum renormalization.
Thus, in addition to calculating the left-hand side, to obtain
$\langle p|{\cal O}_m|p\rangle$ one must also compute the normalization
factor $Z_{nm}$ and cope with the terms of order~$a$.

We know how to calculate $Z_{nm}$, $K_{\sigma F}$, and $K_{nj}$.
They depend on details of the lattice Lagrangian and the lattice 
operator ${O_n}_{\rm lat}$.
One can introduce parameters (call them $c_{\mathrm{SW}}$
and $c_{nj}$) that directly influence $K_{\sigma F}$ and $K_{nj}$.
Thus, one can adjusted $c_{\mathrm{SW}}$ and $c_{nj}$ until 
$K_{\sigma F}\approx0$ and $K_{nj}\approx0$.
This procedure is called Symanzik improvement~\cite{Symanzik:1983dc}.
There are two ways to compute the $Z$s and $K$s.
One is renormalized perturbation theory, which works because the $Z$s 
and $K$s are short-distance quantities.
In this method, uncertainties of order $\alpha_s^l$ remain.
Usually, these days, only one-loop calculations are available, so $l=2$.
Then it is helpful to apply the Brodsky-Lepage-Mackenzie
prescription~\cite{Brodsky:1983gc} to sum up higher-order terms related
to renormalization parts.
The other method is fully non-perturbative~\cite{Sint:2000vc}.
This sounds as if it is exact, but there are uncertainties in the $K$s
of order $a$.
For the matrix element itself, this is just another error of
order~$a^2$, of which there are many.
The first method has been used by the QCDSF
collaboration~\cite{Gockeler:1995wg}, who have rather comprehensive
results for the proton.
The second method has been used by a Zeuthen-Roma~II
collaboration~\cite{Guagnelli:2000sf}, for the first moment of the
pion structure function.

The quenched results from QCDSF do not agree especially well with
phenomenology.
It is, of course, tempting (and reasonable) to blame the quenched 
approximation.
Till now, one could also blame the phenomenological result, which gets 
a significant contribution from the high-$x$ region, where there are 
no experimental data.
As it turns out, neither is the main culprit.
Earlier this year the LHPC and Sesam collaborations finished a 
comparison of quenched and unquenched (well, $n_f=2$) calculations of 
many proton moments~\cite{Dolgov:2002mn}.
To save computer time, these calculations are done with artificially
large light quark mass (for the reasons explained above).
The dependence on the light quark mass for a typical moment is shown in
Fig.~\ref{fig:pdf}(b).
Ref.~\cite{Dolgov:2002mn} finds hardly any difference between unquenched
and quenched calculations for $0.7m_s<m_q<1.6m_s$.
It makes a huge difference, however, whether one follows one's nose and
extrapolates linearly, or whether one follows chiral perturbation
theory.
The latter, of course, is correct.
It has a pronounced curvature for small quarks masses, of the form
$m_\pi^2\ln m_\pi^2$ (and $m_\pi^2\propto m_q$).
With chiral perturbation theory, the extrapolated result agrees with
phenomenology.

\section{Flavor Physics}
\label{sec:flavor}

The central question in flavor physics is whether the standard CKM 
mechanism explains all flavor and $CP$ violation (in the quark 
sector).
One angle on this question is over-constraint of the CKM matrix.
Because the CKM matrix has only four free parameters, the magnitudes 
of the CKM matrix elements dictate the $CP$ violating phase.
Many of the magnitudes may be obtained from
semi-leptonic decays, such as $K\to\pi l\nu$ for the Cabibbo angle, 
$B\to D^*l\nu$ for $|V_{cb}|$, and $B\to\pi l\nu$ for $|V_{ub}|$.
CKM elements on the third row (involving the top quark) enter through
neutral meson mixing, in the neutral $K$, $B$, and $B_s$ systems.
Here we will focus on $B$ physics, with a few remarks on
$K^0$-$\bar{K}^0$ mixing in Sec.~\ref{subsec:kaon}.

For $B$ physics,  one must confront heavy quark discretization effects.
Compared to the lattice spacing, the $b$ quark mass is large, $m_ba>1$.
The Symanzik effective field theory, at least as usually 
applied, breaks down.
Lattice gauge theory does not break down, however, and the Isgur-Wise
heavy-quark symmetries emerge, in the usual way,
for all $m_Qa$~\cite{El-Khadra:1997mp}.
Thus, as long as $m_Q\gg\Lambda_{\rm QCD}$,
lattice gauge theory can be described by heavy-quark effective theory
(HQET).
One can write~\cite{Kronfeld:2000ck}
\begin{equation}
	{\cal L}_{\rm lat} \doteq
		\sum_n {\cal C}_n^{\mathrm{lat}}(m_Q, m_Qa; \mu) {\cal O}_n(\mu),
	\label{eq:HQETlat}
\end{equation}
where $\doteq$ means ``has the same matrix elements as.''
In the same way
\begin{equation}
	{\cal L}_{\rm QCD} \doteq 
		\sum_n {\cal C}_n^{\mathrm{cont}}(m_Q; \mu)      {\cal O}_n(\mu).
	\label{eq:HQETcont}
\end{equation}
The difference is in the short-distance coefficients ${\cal C}_n$.
On the lattice there are two short distances, $a$ and $m_Q^{-1}$, so
the ${\cal C}_n^{\mathrm{lat}}$ depend on the ratio $a/m_Q^{-1}=m_Qa$.
On the other hand, the operators ${\cal O}_n$ in Eqs.~(\ref{eq:HQETlat}) 
and~(\ref{eq:HQETcont}) are essentially the same.

One can therefore systematically improve lattice calculations of
$b$-flavored hadrons, by matching lattice gauge theory and continuum
QCD such that
\begin{equation}
	\delta{\cal C}_n =
		{\cal C}_n^{\mathrm{cont}}-{\cal C}_n^{\mathrm{lat}}
		\approx 0,
\end{equation}
for the first several operators.
HQET is, here, merely an analysis tool; details of how HQET is defined
and renormalized drop out of the difference.
These ideas are put to direct use in the Fermilab 
method~\cite{El-Khadra:1997mp,Kronfeld:2000ck}, which is based on 
Wilson fermions and, thus, also possesses a smooth continuum limit.
Similar ideas are put to use in lattice
NRQCD~\cite{Lepage:1987gg}, which discretizes the continuum heavy-quark
Lagrangian.

\subsection{$B\to D^*l\nu$, ${\cal F}_{B\to D^*}(1)$, and $|V_{cb}|$}

To determine $|V_{cb}|$ from the semi-leptonic decay $B\to D^*l\nu$,
one measures the differential decay rate in $w$, which is the velocity
transfer from the $B$ to the $D^*$.
Then, one extrapolates to zero recoil, $w=1$.
Thus, one can summarize the experiment by saying it measures
$|V_{cb}|{\cal F}_{B\to D^*}(1)$, where
${\cal F}_{B\to D^*}(w)$ is a certain combination of form factors.
At zero recoil all form factors but $h_{A_1}$ are suppressed, so
\begin{equation}
	{\cal F}_{B\to D^*}(1) = h_{A_1}(1) =
		\langle D^*(v) | {\cal A}^\mu | B(v) \rangle.
\end{equation}
It should be ``straightforward'' to calculate this matrix element
in lattice QCD.
But a brute force calculation of $\langle D^*|{\cal A}^\mu|B\rangle$
would not be interesting: similar matrix elements like
$\langle 0|{\cal A}^\mu|B\rangle$ and (see below)
$\langle\pi|{\cal V}^\mu|B\rangle$ have 15--20\% errors.

At zero recoil heavy-quark symmetry constrains $h_{A_1}(1)$ to take the
form
\begin{equation}
	h_{A_1}(1) = \eta_A \left[1+
		\delta_{1/m^2} +
		\delta_{1/m^3} \right] ,
	\label{eq:hA1_anatomy}
\end{equation}
where $\eta_A$ is a short-distance coefficient,
and the $\delta_{1/m^n}$ are (principally)
long-distance matrix elements in HQET at order $1/m^n$.
HQET does not provide the tools to calculate them, but with the 
insight from matching lattice gauge theory to HQET, we have recently 
figured out how to do so~\cite{Hashimoto:2001nb}.
Furthermore, since we incorporate heavy-quark symmetry from the 
outset, and all our uncertainties scale as $h_{A_1}-1$.

From HQET, the structure of the $1/m_Q^n$ corrections is
\begin{eqnarray}
	\delta_{1/m^2} & = &
		- \frac{  \ell_V}{(2m_c)^2}
		+ \frac{ 2\ell_A}{(2m_c)(2m_b)}
		- \frac{  \ell_P}{(2m_b)^2} \label{eq:ell}\\
	\delta_{1/m^3} & = &
		- \frac{  \ell_V^{(3)}}{(2m_c)^3}
		+ \frac{  \ell_A^{(3)}\Sigma +
		\ell_D^{(3)}\Delta}{(2m_c)(2m_b)}
		- \frac{  \ell_P^{(3)}}{(2m_b)^3} \label{eq:ell3}
\end{eqnarray}
where $\Sigma=1/(2m_c)+1/(2m_b)$ and $\Delta=1/(2m_c)-1/(2m_b)$.
In lattice gauge theory, we seek objects whose heavy-quark expansions 
contain the $\ell$s.
From work on the $B\to D$ form factor~\cite{Hashimoto:1999yp}, we 
know certain ratios have small enough uncertainties.
Moreover, one can show via HQET that~\cite{Kronfeld:2000ck}
\begin{eqnarray}
	\frac{\langle D   |\bar{c}\gamma^4 b| B   \rangle
		  \langle B   |\bar{b}\gamma^4 c| D   \rangle}
		 {\langle D   |\bar{c}\gamma^4 c| D   \rangle
		  \langle B   |\bar{b}\gamma^4 b| B   \rangle} & = &
	\left\{ \eta_V^{\rm lat} \left[ 1 - \ell_P\Delta^2 -
		\ell_P^{(3)} \Delta^2\Sigma \right] \right\}^2 ,
	\label{eq:R+} \\
	\frac{\langle D^* |\bar{c}\gamma^4 b| B^* \rangle
		  \langle B^* |\bar{b}\gamma^4 c| D^* \rangle}
		 {\langle D^* |\bar{c}\gamma^4 c| D^* \rangle
		  \langle B^* |\bar{b}\gamma^4 b| B^* \rangle} & = &
	\left\{ \eta_V^{\rm lat} \left[ 1 - \ell_V \Delta^2 -
		\ell_V^{(3)} \Delta^2\Sigma \right] \right\}^2 ,
	\label{eq:R1} \\
	\frac{\langle D^* |\bar{c}\gamma^j \gamma_5 b| B   \rangle
		  \langle B^* |\bar{b}\gamma^j \gamma_5 c| D   \rangle}
		 {\langle D^* |\bar{c}\gamma^j \gamma_5 c| D   \rangle
		  \langle B^* |\bar{b}\gamma^j \gamma_5 b| B   \rangle} & = &
	\left\{ \check{\eta}_A^{\rm lat} \left[ 1 - \ell_A\Delta^2 -
		\ell_A^{(3)} \Delta^2\Sigma \right] \right\}^2 ,
	\label{eq:RA1} 
\end{eqnarray}
and one-loop expansions of $\eta_V^{\rm lat}$ and 
$\check{\eta}_A^{\rm lat}$ are available~\cite{Harada:2001fj}.
By calculating the ratios for many combinations of the heavy-quark
masses, we can fit to the HQET description on the right hand side
to obtain all three $\ell$s in $\delta_{1/m^2}$, and three of four
$\ell^{(3)}$s in~$\delta_{1/m^3}$.
We can then reconstitute $h_{A_1}(1)$ with Eq.~(\ref{eq:hA1_anatomy}),
finding
\begin{equation}
	{\cal F}_{B\to D^*}(1) = h_{A_1}(1) = 0.913 
		{}^{+0.024}_{-0.017} 
		\pm 0.016 
		{}^{+0.003}_{-0.014} 
		{}^{+0.000}_{-0.016} 
		{}^{+0.006}_{-0.014} ,
	\label{eq:F(1)}
\end{equation}
where the uncertainties stem from statistics and fitting, HQET matching,
lattice spacing dependence, the chiral extrapolation, and the effect
of the quenched approximation.
Instead of adding these errors in quadrature, we prefer to take note of
a bound, ${\cal F}_{B\to D^*}(1)\leq1$, and posit a Poisson distribution
$P(x) = N x^7 e^{-7x}$, $x = [1-{\cal F}_{B\to D^*}(1)]/0.087\geq0$,
for global fits of the CKM matrix~\cite{Kronfeld:2002cc}.

\subsection{$B\to\pi l\nu$, $f_+(E)$, and $|V_{ub}|$}

To determine $|V_{ub}|$ from the semi-leptonic decay $B\to\pi l\nu$,
it is not just a rehash of the previous section.
The experimental rate is smaller, by a factor $|V_{ub}/V_{cb}|^2$,
and heavy-quark symmetry is not as constraining.
Experiments should measure~\cite{Warburton}
\begin{equation}
	\int_{E_l^{\rm min}}^{E_l^{\rm max}}dE_l\,
	\int_{E_\pi^{\rm min}}^{E_\pi^{\rm max}}dE_\pi\,
		\frac{d^2\Gamma}{dE_ldE_\pi} \propto |V_{ub}|^2 
	\int_{E_\pi^{\rm min}}^{E_\pi^{\rm max}}dE_\pi\,p^3
		|f_+(E_\pi)|^2,\quad
\end{equation}
where $E_\pi=v\cdot p_\pi$ is pion energy in the $B$ rest frame,
$p^2=E_\pi^2-m_\pi^2$, and $E_l$ is charged lepton energy.
To determine $|V_{ub}|$ one needs a reliable calculation 
of the form factor~$f_+(E)$, which parametrizes the matrix element of 
the $b\to u$ vector current.
Any cut on the lepton variable is equally good~\cite{Warburton}.

Recently there have been several calculations of these form factors,
using several different methods~\cite{Bowler:1999xn,Abada:2000ty,%
El-Khadra:2001rv,Aoki:2001rd,Shigemitsu:2002wh}.
Two of these works~\cite{Bowler:1999xn,Abada:2000ty} calculate the
matrix element with $m_Q$ around the charm mass,
fit to a model for the $E_\pi$ dependence, and extrapolate the model
parameters with heavy-quark scaling.
The others appeal more directly to heavy-quark ideas, as discussed
above.
El-Khadra \emph{et al.}~\cite{El-Khadra:2001rv} use the Fermilab
method, and the other two~\cite{Aoki:2001rd,Shigemitsu:2002wh} use
lattice NRQCD.
Refs.~\cite{El-Khadra:2001rv,Aoki:2001rd,Shigemitsu:2002wh} do not
use a model for the $E_\pi$ dependence; instead a cut on~$E_\pi$
is used to control discretization effects.

An obvious challenge in these calculations arises from
discretization errors of the final-state pion,
which grow as $\bbox{p}_\pi a$.
This makes it hard to get $E_\pi=2.6~{\rm GeV}$.
A possibility to circumvent this difficulty is to give $B$ meson
momentum~\cite{Foley}.
For example, if one chooses $-\bbox{p}_B=\bbox{p}_\pi=800~{\rm MeV}$
in the lattice frame of reference, one can access the whole kinematic
range.
A less obvious, but also important, challenge is the 
chiral extrapolation.
It is not well understood and contributes the largest
systematic error in the calculation with the smallest quark
masses~\cite{El-Khadra:2001rv}.
The uncertainties on $f_+(E)$ are still 15--20\% in the quenched
approximation.
But there are no real technical roadblocks (for details, see 
Ref.~\cite{El-Khadra:2001rv}), so the errors will be reduced while 
BaBar and Belle accumulate data.
In the short term, it will be interesting and important to compare 
similar lattice calculations for semi-leptonic $D$ decays to 
experimental results from CLEO-$c$.

\subsection{Neutral $B$ Mixing and $|V_{td}|$}

In the Standard Model, neutral meson mixing is sensitive to $V_{td}$ 
and~$V_{ts}$.
A significant recent development is the realization that the 
theoretical uncertainty in $B^0$-$\bar{B}^0$ mixing has been 
underestimated.
The culprit has been the chiral extrapolation, which we have seen to 
be important in moments of structure functions.

In the Standard Model, the oscillation frequency for
$B^0_d$-$\bar{B}^0_d$ mixing is
\begin{equation}
	\Delta m_d \propto |V_{td}|^2 {\cal M}_d
\end{equation}
where ${\cal M}_q=%
\langle\bar{B}^0_q|[\bar{b}(1-\gamma_\mu)\gamma_5q]%
[\bar{b}(1-\gamma_\mu)\gamma_5q]|B^0_q\rangle$.
Phenomenologists usually write 
${\cal M}_q=\frac{8}{3}m_{B_q}^2f_{B_q}^2B_{B_q}$ but lattice 
calculations give matrix elements ${\cal M}_q$ directly, and 
$f_{B_q}$ from $\langle0|\bar{b}\gamma_\mu\gamma_5q|B^0_q\rangle$.
Nevertheless, it turns out to be useful to look separately at
$f_{B_q}$ and $B_{B_q}$.
Current world averages (from lattice QCD) are
$f_{B_q} = 198 \pm 30~{\rm MeV}$ and
$B_{B_q} = 1.30 \pm 0.12$~\cite{Ryan:2001ej}.
So the error on $|V_{td}|$ from $\Delta m_d$ alone is $\sim15\%$.

For some time, the conventional wisdom has said that most of the
theoretical uncertainty cancels if one takes the
ratio~$\Delta m_s/\Delta m_d$.
(It is anticipated that $\Delta m_s$ will be measured at Run~2 of the 
Tevatron~\cite{Anikeev:2001rk}.)
The ratio is
\begin{equation}
	\frac{\Delta m_s}{\Delta m_d} =
		\left|\frac{V_{ts}}{V_{td}}\right|^2
		\frac{m_{B_s}}{m_{B_d}}
		\xi^2, \quad
	\xi^2 = \frac{f^2_{B_s}B_{B_s}}{f^2_{B_d}B_{B_d}}.
\end{equation}
CKM unitarity says $|V_{ts}|=|V_{cb}|$ to good approximation, 
and $|V_{cb}|$ is known to 2--4\%.
Many authors believe the uncertainty in~$\xi$ to be less than~5\%.
Cancellations do occur in the statistical error, and in systematics
at short distances ($a$ and $m_b^{-1}$) and---arguably---at medium
distances ($\Lambda_{\rm QCD}^{-1}$).
But they explicitly do \emph{not} cancel at long distances between
$m_s^{-1}$ and $m_d^{-1}$ from light quarks in the $B_s$ and $B$~mesons.

Now, the quenched approximation does not work well at these long
distances, and unquenched calculations are prohibitive at $m_q\sim m_d$.
Thus, $\xi$ isolates the contributions that are hardest to capture,
and tries to get at them by extrapolating in~$m_q$.
After studying the differences in chiral logarithms in real QCD and
the quenched approximation, Booth~\cite{Booth:1994hx} and Sharpe and
Zhang~\cite{Sharpe:1996qp} sounded notes of caution.
Their analyses showed that chiral logarithms should induce curvature as
a function of light quark mass, which quenching would obscure.
This curvature has recently been observed in unquenched calculations
(well, $n_f=2$ again), and identified as a serious source of
uncertainty~\cite{Yamada:2001xp}.

\begin{figure}[b!p]
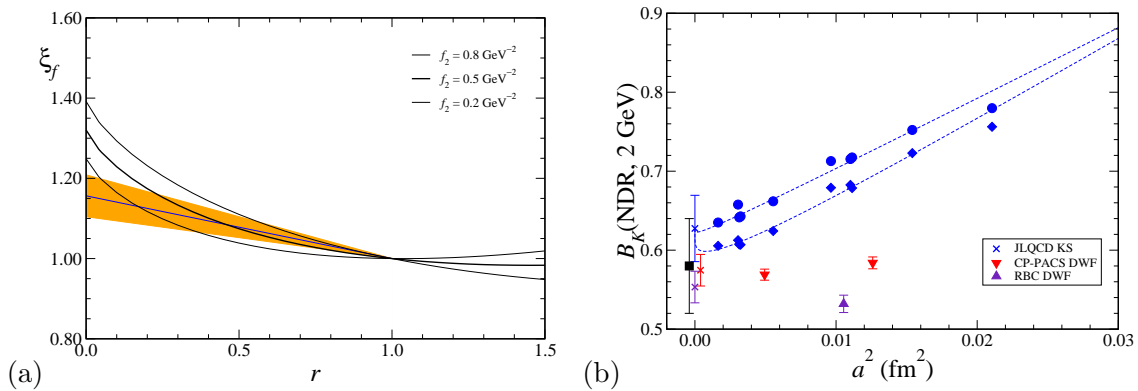

	{\small (a)}\hspace*{-0.5em}
	\includegraphics[width=0.46\textwidth]{xi_from_f2.eps}\hfill
	{\small (b)}\hspace*{-0.5em}
	\includegraphics[width=0.46\textwidth]{BK.eps}
	\caption[Neutral meson mixing]{\it Neutral meson mixing.
	(a)~Comparison of chiral extrapolations
	of~$\xi$~\cite{Kronfeld:2002ab}.
	(b)~$B_K$ vs.\ $a^2$, with results from
	JLQCD~\cite{Aoki:1997nr},
	CP-PACS~\cite{AliKhan:2001wr}, and
	RBC~\cite{Blum:2001xb}.}
	\label{fig:mix}
\end{figure}

It is not too difficult to grasp the problem.
For convenience, let $\xi = \xi_f \xi_B$, where
$\xi_f = f_{B_s}/f_{B_d}$, $\xi_B^2 = B_{B_s}/B_{B_d}$.
Chiral perturbation theory says that
\begin{equation}
	\xi_f(r) - 1 = m_{ss}^2 (1 - r) \left[\case{1}{2} f_2 -
		\frac{1+3g_{BB^*\pi}^2}{(4\pi f_\pi)^2} l(r) \right],
	\label{eq:xi_chi_log}
\end{equation}
where $m_{ss}^2=2m_K^2-m_\pi^2$,
$r=m_q/m_s$ measures the light quark mass in units of the strange mass,
$f_\pi$ is the pion decay constant, and
$g_{BB^*\pi}$ is the $B$-$B^*$-$\pi$ coupling.
The function $l(r)$ contains chiral logarithms:
\begin{equation}
\begin{array}{rccccl}
	(1-r)l(r) \; = \; &
		\frac{1}{4} (1+r)\ln\left[(1+r)/2\right] & + &
		\frac{1}{12}(2+r)\ln\left[(2+r)/3\right] & - &
		\frac{3}{4}   r  \ln(r) \\[0.7em]
		B_s~\textrm{mixing:} \quad
		& B_s\leftrightarrow B^*K      & &
		  B_s\leftrightarrow B^*_s\eta & & \\
		B_d~\textrm{mixing:} \quad
		& B  \leftrightarrow B_s^*K    & &
		  B  \leftrightarrow B^*\eta   & &
		  B  \leftrightarrow B^*\pi 
\end{array}
\end{equation}
and each term arises from the virtual corrections given beneath it.
All other contributions are described well enough by linear behavior
in~$r$ and are lumped into the constant $f_2$.
The ratio $\xi^2_B$ is described by an expression similar
to Eq.~(\ref{eq:xi_chi_log}), except that the chiral log is multiplied
by~$1-g_{BB^*\pi}^2$.
Unquenched lattice calculations~\cite{Yamada:2001xp,Bernard:2002pc}
are not yet good enough to determine directly the coefficients of the
chiral logs.
Sin\'ead Ryan and I have suggested taking them from phenomenology
instead~\cite{Kronfeld:2002ab}.
We invoke heavy-quark symmetry, which says
the $B$-$B^*$-$\pi$ coupling should be roughly the same as the 
$D$-$D^*$-$\pi$ coupling.
Then the recent measurement of the $D^*$ width yields 
$g_{DD^*\pi}^2=0.35$~\cite{Anastassov:2001cw};
we take $g_{BB^*\pi}^2=0.35\pm20\%$.
We obtain the constant $f_2$ from the slope of $\xi_f(r)$
around $0.5<r<1.0$, where quenched and unquenched calculations are in
good agreement.
We also analyze $\xi^2_B$ in the same manner.
Finally, we find
\begin{equation}
	\xi = 1.32 \pm 0.10 \quad \textrm{(chiral log extrapolation)}
\end{equation}
from the chiral log fit.
Previously, one had tried linear fits, which would have given
\begin{equation}
	\xi = 1.15 \pm 0.05 \quad \textrm{(conventional linear extrapolation)}
	\label{eq:xi_old}
\end{equation}
for the same input.
The difference is illustrated in Fig.~\ref{fig:mix}(a).
One sees that the uncertainty of 5\% [as in Eq.~(\ref{eq:xi_old})]
certainly was underestimated, and also that the central value is
probably quite different from the conventional 1.15.

\subsection{Kaon Physics}
\label{subsec:kaon}

Two theoretical developments of the past few years have opened the
way for a wider range of kaon calculations.
One is a method for exploiting finite-volume effects to calculate
phase shifts in $K\to\pi\pi$ (and other quasi-elastic
processes)~\cite{Lellouch:2000pv}.
The other is the formulation of lattice fermions with good chiral
symmetry~\cite{Shamir:1993zy,Wiese:1993cb,Neuberger:1997fp}.
With these new tools, we may be able to obtain, at last, quantitative 
results for such long-standing problems as the $\Delta I=\case{1}{2}$ 
rule and the matrix elements needed to compute 
$\varepsilon'/\varepsilon$ in the Standard Model.

Ref.~\cite{Lellouch:2000pv} (and earlier papers by L\"uscher) base a
formalism for calculating final-state phases on three insights.
The first is that energy levels in finite volumes are discrete.
The second is that phase shifts arise at hadronic distances of order 
1~fm, remote from the box size $L>1$~fm.
Lastly, there is a kinematical, albeit very non-trivial, set of 
$L$-dependent relationships between the phase shifts and the energy 
levels.
As a result, one can extract the (elastic) phase shifts from the $L$ 
dependence of the discrete energy spectrum.
Unfortunately, practical application to QCD requires box sizes 
$L=2$--$6~{\rm fm}$, so it will not be feasible soon.

In many kaonic matrix elements, chiral symmetry is important, for 
example for maintaining a simple relation between lattice operators 
and their counterparts in continuum QCD.
In 1982 Ginsparg and Wilson derived a sufficient condition for chiral
symmetry on the lattice~\cite{Ginsparg:1981bj}.
For a long time the Ginsparg-Wilson relation defied solution, buy now
there are at least two realizations,
the ``fixed-point action''~\cite{Wiese:1993cb},
and ``overlap fermions''~\cite{Neuberger:1997fp}.
Another method, ``domain-wall fermions''~\cite{Shamir:1993zy},
is related to the former and has exponentially small violations of
the Ginsparg-Wilson relation.
Two lattice collaborations have large-scale calculations of~$B_K$
(defined analogously to $B_B$) with domain-wall quarks.
In Fig.~\ref{fig:mix}, published results from
CP-PACS~\cite{AliKhan:2001wr} and RBC~\cite{Blum:2001xb} are compared to
classic work of JLQCD (with Kogut-Susskind quarks)~\cite{Aoki:1997nr}.
The lattice spacing dependence seems gentler for domain-wall 
fermions.
From looking at the plot, a rough estimate of an average would be
\begin{equation}
	B_K({\rm NDR}, 2~{\rm GeV}) = 0.58 \pm 0.06
\end{equation}
which encompasses also Ref.~\cite{Kilcup:1997ye}.
This uncertainty could easily be reduced, by using, say, 3--5 lattice
spacings with domain-wall (or overlap) fermions, and taking the
continuum limit.
One should also keep in mind that these calculations have been done in
the quenched approximation, and with degenerate quarks of mass~$m_s/2$.

\section{Conclusions and Prospects}
\label{sec:end}

Although the foundation of lattice QCD is sound, some difficulties
arise when turning the idealized theory into a computation tool.
Errors are introduced at short and long distances.
They are controlled by effective field theories, however, providing
reliable methods to obtain physical predictions.
A~wide variety of calculations in the quenched approximation have
allowed us to learn how control short-distance effects:
both for light quarks and for heavy quarks.
Now that several (partially) unquenched calculations are available,
other issues are becoming clearer, particularly the chiral
extrapolation of light quark masses.
One can be optimistic that unquenched calculations---with solid,
transparent analyses of all uncertainties---will become available to
help interpret experiments with high-energy collisions.

One upcoming program is especially noteworthy vis a vis lattice QCD.
In the next few years, CLEO-$c$~\cite{Briere:2001rn} will measure
leptonic and semi-leptonic decays of $D$ and $D_s$ mesons to a few
per~cent.
Lattice QCD has a chance to predict their results, perhaps with
comparable accuracy.
An especially interesting combination is $f^{D\to Kl\nu}_+(E)/f_{D_{s}}$
(and the Cabibbo-suppressed cousin $f^{D\to\pi l\nu}_+(E)/f_D$).
The CKM matrix drops out from the measurements, and non-Standard
physics is unlikely.
Thus, one has direct tests of non-perturbative QCD.
The ratio $f_{D_s}/f_D$ is also interesting, because it tests the
chiral extrapolation of~$\xi$ in $B$-$\bar{B}$ mixing.
Successful comparisons of lattice calculations and CLEO-$c$ will give
confidence in other applications of lattice QCD, such as $B$ physics
and moments of the parton densities.

\section*{Acknowledgments}
Fermilab is operated by Universities Research Association Inc.,
under contract with the U.S.\ Department of Energy.

\end{document}